\documentclass[conference]{IEEEtran}
\IEEEoverridecommandlockouts
\usepackage{cite}
\usepackage{amsmath,amssymb,amsfonts}
\usepackage{algorithm}
\usepackage{algpseudocode}
\usepackage{amsmath}
\usepackage{graphicx}
\usepackage{textcomp}
\usepackage{xcolor}
\usepackage{booktabs}
\usepackage{multirow}
\usepackage{array}
\def\BibTeX{{\rm B\kern-.05em{\sc i\kern-.025em b}\kern-.08em
    T\kern-.1667em\lower.7ex\hbox{E}\kern-.125emX}}

\usepackage{acronym}
\newacro{LLMs}{Large Language Models}
\newacro{LLM}{Large Language Model}
\newacro{K}{Key}
\newacro{V}{Value}
\newacro{KV} {Key-Value}
\newacro{Q}{Query}
\newacro{O}{Attention Output}
\newacro{QO}{Query and Attention Output}
\newacro{eDRAM}{embedded DRAM}
\newacro{CNN}{Convolutional Neural Networks}
\newacro{SHIELD}{Segmented Hierarchical Inference with Ephemeral Lifecycle Design}
\newacro{NPUs}{Neural Processing Units}
\newacro{NPU}{Neural Processing Unit}
\newacro{BER}{Bit Error Rate}
\newacro{FFN}{Feed-Forward Networks}
\newacro{CNN}{Convolutional Neural Network}
\newacro{DNN}{Deep Neural Network}
\newacro{BF16}{bfloat16}

\algdef{SE}[FUNCTION]{Function}{EndFunction}[2]{\algorithmicfunction\ \textsc{#1}(#2)}{\textbf{end function}}

\begin{document}

\title{
SHIELD: A Segmented Hierarchical Memory Architecture for Energy-Efficient LLM Inference on Edge NPUs
}

\author{\IEEEauthorblockN{Jintao Zhang, Xuanyao Fong}
\IEEEauthorblockA{Department of Electrical and Computer Engineering, National University of Singapore, Singapore\\
Email: zhangjintao@u.nus.edu, kelvin.xy.fong@nus.edu.sg}}

\maketitle

\begin{abstract}
\ac{LLM} inference on edge \ac{NPUs} is fundamentally constrained by limited on-chip memory capacity. Although high-density \ac{eDRAM} is attractive for storing activation workspaces, its periodic refresh consumes substantial energy.
Prior work has primarily focused on reducing off-chip traffic or optimizing refresh for persistent \ac{KV} caches, while transient and error-resilient \ac{QO} activations are largely overlooked.
We propose SHIELD, a lifecycle-aware segmented \ac{eDRAM} architecture that jointly exploits temporal residency and bit-level sensitivity in \ac{BF16} activations.
SHIELD isolates the sign and exponent fields from the mantissa, disables refresh for transient QO mantissas, and applies relaxed refresh to persistent KV mantissas.
Across multiple LLMs and inference scenarios, SHIELD reduces \ac{eDRAM} refresh energy by 35\% relative to a standard-refresh baseline while preserving accuracy on WikiText-2, PIQA, and ARC-Easy.
\end{abstract}


\section{Introduction}
The ``Memory Wall'' is a central obstacle to deploying LLMs on edge NPUs, where intermediate activations often exceed the limited on-chip SRAM capacity~\cite{zheng2025review}. 
A practical solution is to place these activations in higher-density on-chip eDRAM, but the required periodic refresh can consume a substantial fraction of memory energy~\cite{xia2025kelle}.
Among these activations, the attention workspace is particularly costly because each layer produces multiple large \(N \times d\) tensors, including \ac{Q}, \ac{K}, \ac{V}, and \ac{O}, defined as follows~\cite{vaswani2017attention}:
\begin{IEEEeqnarray}{rCl}
Q_i, K_i, V_i & = & XW_i^Q, XW_i^K, XW_i^V; \\
H_i & = & \text{Softmax}\left(\frac{Q_i K_i^T}{\sqrt{d_k}}\right)V_i; \\
O & = & \operatorname{Concat}[H_i]_{i=1}^m
\end{IEEEeqnarray}
where \(m\) is the number of attention heads. 
For example, in \textit{Qwen3-8B}~\cite{yang2025qwen3}, the transient \ac{QO} workspace alone requires about 32~MB, far beyond the 1 to 8~MB SRAM budget of representative edge NPUs such as \textit{AMD XDNA}~\cite{2024amd}.
These intermediates must therefore be placed in eDRAM, making refresh energy a first-order design concern.
Existing approaches only partially address this problem (see Table~\ref{tab_related_work}).
RANA~\cite{8416839} optimizes refresh for \ac{CNN} and \ac{DNN} workloads but does not provide the granularity required for LLM inference.
FlashAttention~\cite{dao2022flashattention} and LEAP~\cite{wang2025leap} reduce off-chip I/O through tiling and do not address the refresh energy of \ac{eDRAM}.
Kelle~\cite{xia2025kelle} is the closest prior work, as it optimizes eDRAM refresh for the persistent KV cache; however, it does not address the substantial refresh cost of transient QO activations.
To address these limitations, we propose \textbf{SHIELD}: \textbf{S}egmented \textbf{H}ierarchical \textbf{I}nference with \textbf{E}phemeral \textbf{L}ifecycle \textbf{D}esign, which jointly exploits lifecycle asymmetry and bit-level error tolerance across the activation workspace.
Our contributions are:
\begin{itemize}
\item Segmented Data Mapping: Partitioning activations by lifecycle (QO vs. KV) and numerical significance (exponent vs. mantissa).
\item Hierarchical Refresh Control: Assigning refresh intervals according to each segment’s error tolerance and residency time.
\item Refresh-less Workspace: Eliminating refresh for QO mantissas while applying a relaxed refresh ($T_{rel}=1216\mu s$) to persistent \ac{KV} mantissas.
\end{itemize}

\begin{table}[t]
\caption{Comparison of prior memory-optimization methods for AI accelerators.}
\label{tab_related_work}
\centering
\footnotesize
\setlength{\tabcolsep}{2.5pt}
\renewcommand{\arraystretch}{0.95}
\begin{tabular}{>{\centering\arraybackslash}p{0.19\columnwidth}
                >{\centering\arraybackslash}p{0.20\columnwidth}
                >{\centering\arraybackslash}p{0.14\columnwidth}
                >{\centering\arraybackslash}p{0.33\columnwidth}}
\hline
\textbf{Work} & \textbf{Granularity} & \textbf{Domain} & \textbf{Target} \\
\hline
RANA~[2018]          & Layer / Row  & CNN/DNN & Refresh (short-lived) \\
FlashAttn~[2022]     & Block        & LLM     & Access (HBM I/O) \\
LEAP~[2025]          & Block / Tile & LLM     & Access (PIM NoC) \\
Kelle~[2025]         & 2D (col/row) & LLM     & Refresh (KV only) \\
\textbf{SHIELD}      & \textbf{Segmented} & \textbf{LLM} & \textbf{Refresh (QO + KV)} \\
\hline
\end{tabular}
\vspace{2pt}
\end{table}

\section{Background and Motivation}

\subsection{LLM Activation Lifecycle}
In transformer-based LLM inference, activation tensors exhibit distinct temporal behaviors.
The \ac{QO} matrices are strictly limited to intra-layer dependencies.
In contrast, the \ac{KV} matrices are persistent and must be maintained across the entire context window to support autoregressive generation.

\begin{figure}[t]
    \centering
    \includegraphics[width=\linewidth]{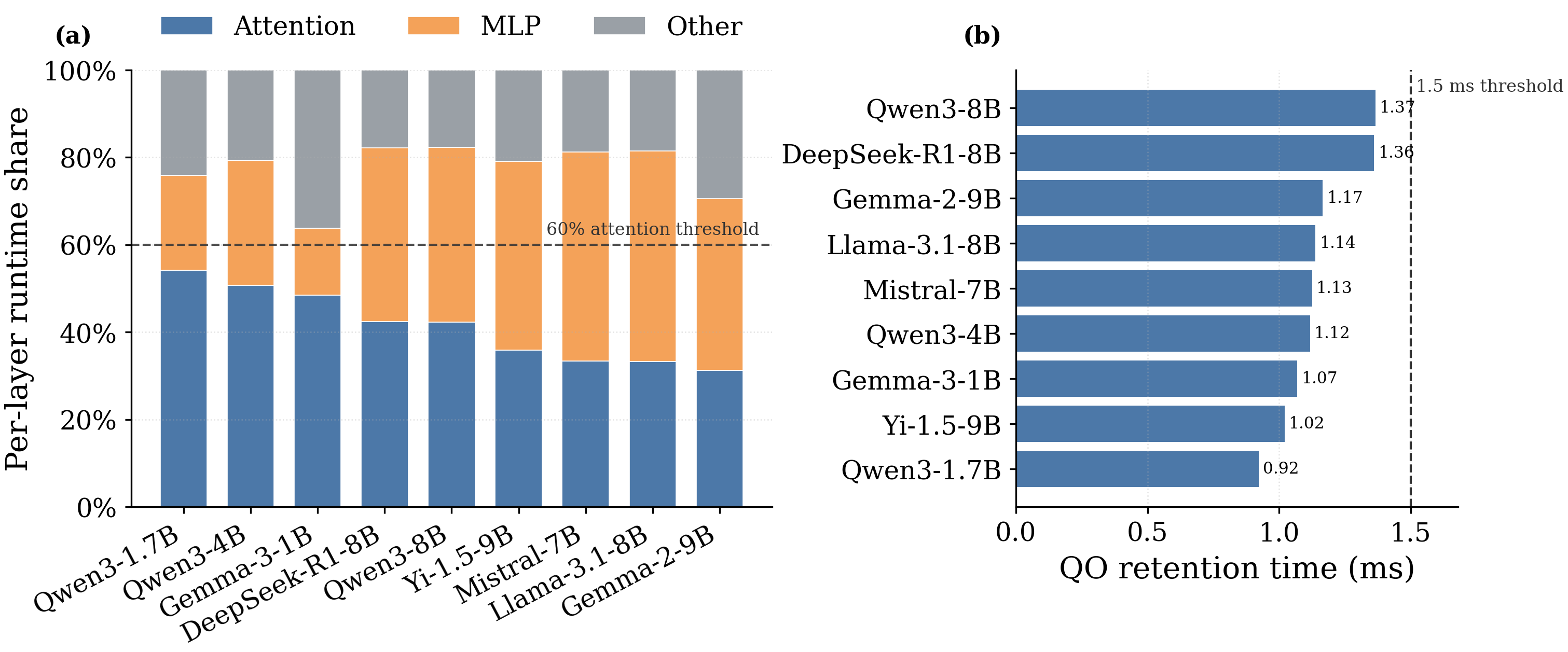}
    \caption{Lifecycle characterization of transient QO activations. (Left) Per-layer runtime composition across evaluated LLMs, showing that attention occupies a limited fraction of execution time. (Right) Measured QO retention time across models remains below 1.5~ms, supporting refresh-less storage for transient mantissas.}
    \label{fig:lifecycle_combined}
\end{figure}

To characterize activation residency, we performed a systematic profiling across a suite of open-source models, including \textit{Qwen}~\cite{yang2025qwen3}, \textit{Mistral}~\cite{jiang2023mistral7b}, and \textit{Llama}~\cite{grattafiori2024llama} families across varying parameter scales (1B to 9B).
As illustrated in Fig.~\ref{fig:lifecycle_combined}(a), our profiling results indicate that the self-attention mechanism consistently accounts for less than 60\% of the total per-layer execution time.
The remaining cycles are consumed by \ac{FFN} and normalization layers, during which the QO buffers remain idle.
This implies that for a large portion of the inference duty cycle, maintaining QO data through active refresh is energy-inefficient.

The feasibility of a refresh-less policy depends on whether the required residency time of activations is shorter than the eDRAM retention time.
We performed a layer retention analysis using a storytelling task with a context length of 2048 on a single NVIDIA H100 GPU.
As shown in Fig.~\ref{fig:lifecycle_combined}(b), the intra-layer processing time and thus, the effective lifetime of QO activations, remains within 1.5~ms across all evaluated models.
These findings establish the rationale for using a refresh-less workspace for transient mantissas.

\subsection{Error Tolerance Heterogeneity}
To evaluate the feasibility of lifecycle-aware refresh gating, we conduct fault-injection experiments with Algorithm~\ref{alg:shield_sim} on BF16 activation storage. The results reveal heterogeneity along two dimensions: bit-level sensitivity and tensor-level resilience.

\begin{figure}[t]
    \centering
    \includegraphics[width=\linewidth]{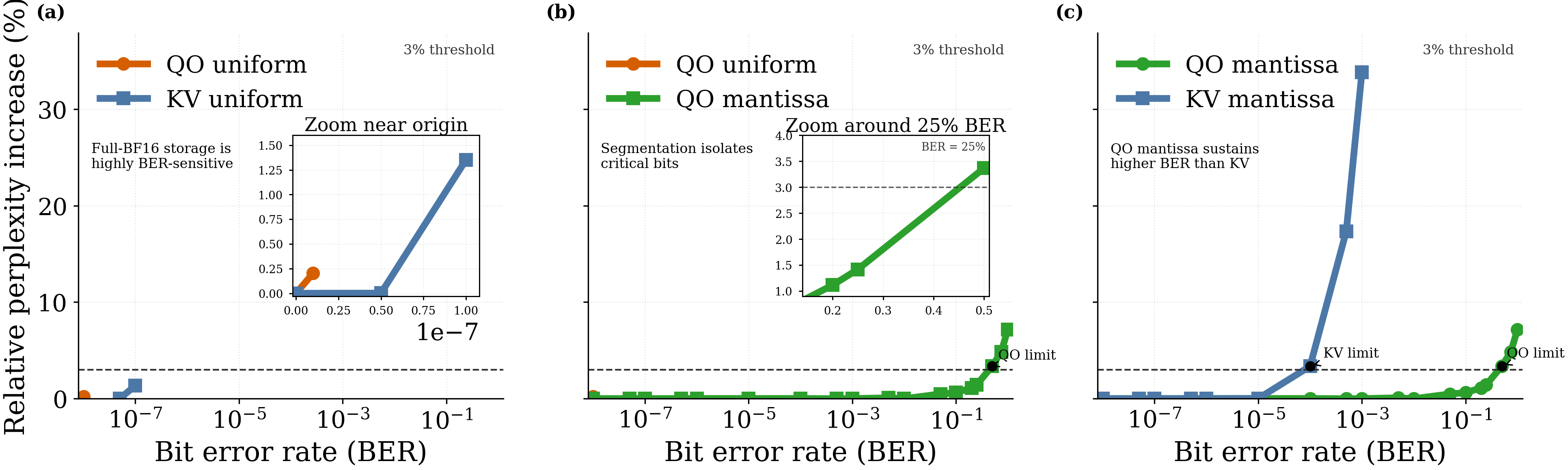}
    \caption{Fault-injection characterization for BF16 activation storage. (Left) Uniform bit errors rapidly degrade perplexity, motivating bit-level segmentation. (Center) Transient QO mantissas tolerate high BER, enabling refresh-less storage. (Right) Persistent KV mantissas require a stricter BER bound for stable long-context generation, motivating relaxed refresh instead of refresh elimination.}
    \label{fig:error_combined}
\end{figure}

When errors are injected uniformly across the QO workspace as shown in Fig.~\ref{fig:error_combined}(a), model perplexity degrades rapidly once the \ac{BER} exceeds $10^{-6}$, indicating that even transient activations remain highly sensitive to corruption in critical bit fields.
This observation motivates a segmented design in which the sign and exponent fields are physically isolated from the more error-tolerant mantissa. 

After isolating the critical fields, we evaluate the resilience of the 7-bit mantissa for transient QO data as shown in Fig.~\ref{fig:error_combined}(b).
Specifically, we define the threshold for numerical instability as a 3\% degradation in the perplexity score compared to the baseline at a 0\% error rate.
The results show that transient QO mantissas tolerate substantially higher error rates, with stable output quality maintained even at an injected error rate of up to 25\%.
Since the measured QO lifetime remains within 1.5~ms, this tolerance supports a refresh-less policy for QO mantissa banks.

In contrast, persistent KV cache mantissas are more sensitive to cumulative error propagation.
Fig.~\ref{fig:error_combined}(c) highlights the stricter BER requirement for persistent KV mantissas.
We observe that the KV mantissa tolerates BERs up to $10^{-4}$ before long-context generation degrades noticeably.
Accordingly, SHIELD employs relaxed refresh rather than refresh elimination for persistent KV mantissa.

\begin{algorithm}[t]
\caption{SHIELD Lifecycle-Aware Fault Simulation}
\label{alg:shield_sim}
\begin{algorithmic}[1]
\small
\Require LLM $\mathcal{M}$, BER $(\rho_{kv}, \rho_{qo})$, Mantissa Mask $\mathbf{M}$
\For{each layer $L \in \mathcal{M}$}
    \State Register \textit{ForwardHook}($L$) $\rightarrow$ \textbf{Output} $\mathbf{X}$
    \State $\rho \leftarrow (L \in \{W_K, W_V\}) ? \rho_{kv} : (L \in \{W_Q, W_O\}) ? \rho_{qo} : 0$
    \If{$\rho > 0$}
        \State $\mathbf{B}_{flip} \leftarrow \text{RandomUniform}(\text{shape}(\mathbf{X})) < \rho$
        \State $\mathbf{N}_{bits} \leftarrow \text{RandomInt16}() \text{ AND } \mathbf{M}$
        \State $\mathbf{X}_{noisy} \leftarrow \text{Reinterpret}(\mathbf{X}) \oplus (\mathbf{N}_{bits} \cdot \mathbf{B}_{flip})$
        \State \textbf{return} \text{ReinterpretAsBF16}($\mathbf{X}_{noisy}$)
    \EndIf
\EndFor
\State \Return $\text{LMEval}(\mathcal{M}, \text{Tasks})$
\end{algorithmic}
\end{algorithm}

\section{SHIELD Architecture}

\begin{figure}[t]
    \centering
    \includegraphics[width=\linewidth]{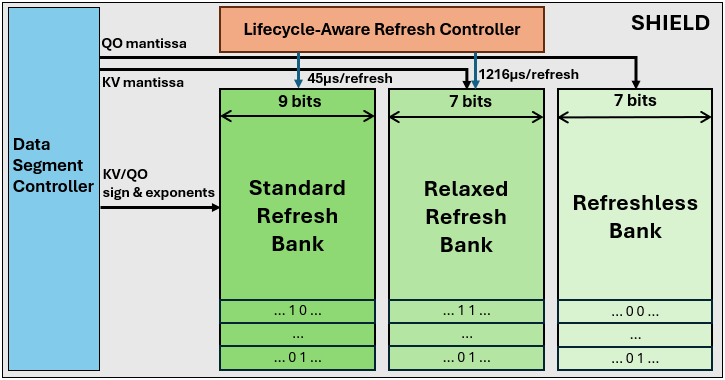}
    \caption{Overview of the SHIELD Architecture, showcasing the segmented memory controller and the mapping of BF16 activations to heterogeneous eDRAM banks.}
    \label{fig:shield_arch}
\end{figure}

The SHIELD architecture is designed to bridge the gap between high-density activation storage and power efficiency.
As illustrated in Fig.~\ref{fig:shield_arch}, the system consists of a segmented memory controller that orchestrates data placement across heterogeneous eDRAM banks based on activation lifecycle and bit-level significance.

\subsection{Segmented Memory Hierarchy}
SHIELD organizes the eDRAM workspace into three banks according to numerical sensitivity and data lifetime in order to support the BF16 format. The 1-bit sign and 8-bit exponent are stored in Standard Refresh Banks, while the 7-bit mantissa is divided by residency: persistent KV cache mantissas are mapped to Relaxed Refresh Banks, and transient QO mantissas are mapped to Refresh-less Banks. Standard and Relaxed Refresh Banks are governed by a lifecycle-aware refresh controller, whereas Refresh-less Banks are operated without refresh.

\subsection{Lifecycle-Aware Refresh Policy}
SHIELD derives its efficiency from the dependence of BER on charge-retention time. To obtain the safe operating region of each memory bank, we conducted empirical characterization on a 3T-eDRAM cell model~\cite{4700556}. Our results indicate that keeping the BER of the 7-bit KV mantissa field below $10^{-4}$ requires a refresh interval of $T_{rel}=1216\,\mu s$. By comparison, the BER of the 7-bit transient QO mantissa field remains bounded at $0.04\%$ over its effective lifetime. These observations provide the empirical basis for the refresh configuration adopted in SHIELD and for the evaluation that follows.

\begin{figure}[t]
    \centering
    \includegraphics[width=\linewidth]{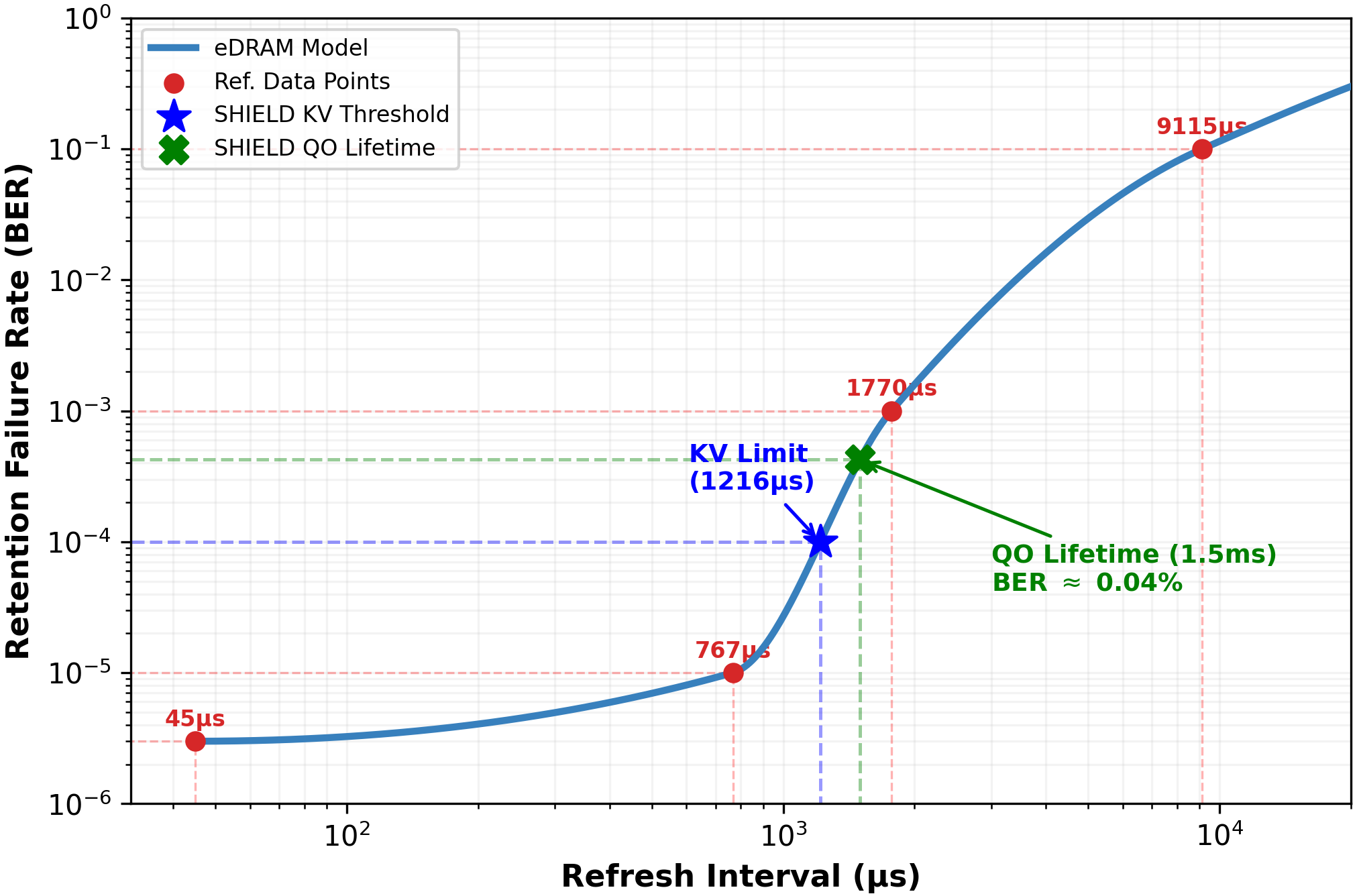}
    \caption{Characterization of 3T-eDRAM BER vs. Retention Time. The plot identifies the optimal relaxation point for the KV cache mantissas.}
    \label{fig:edram_ber}
\end{figure}

\begin{figure}[t]
    \centering
    \includegraphics[width=\linewidth]{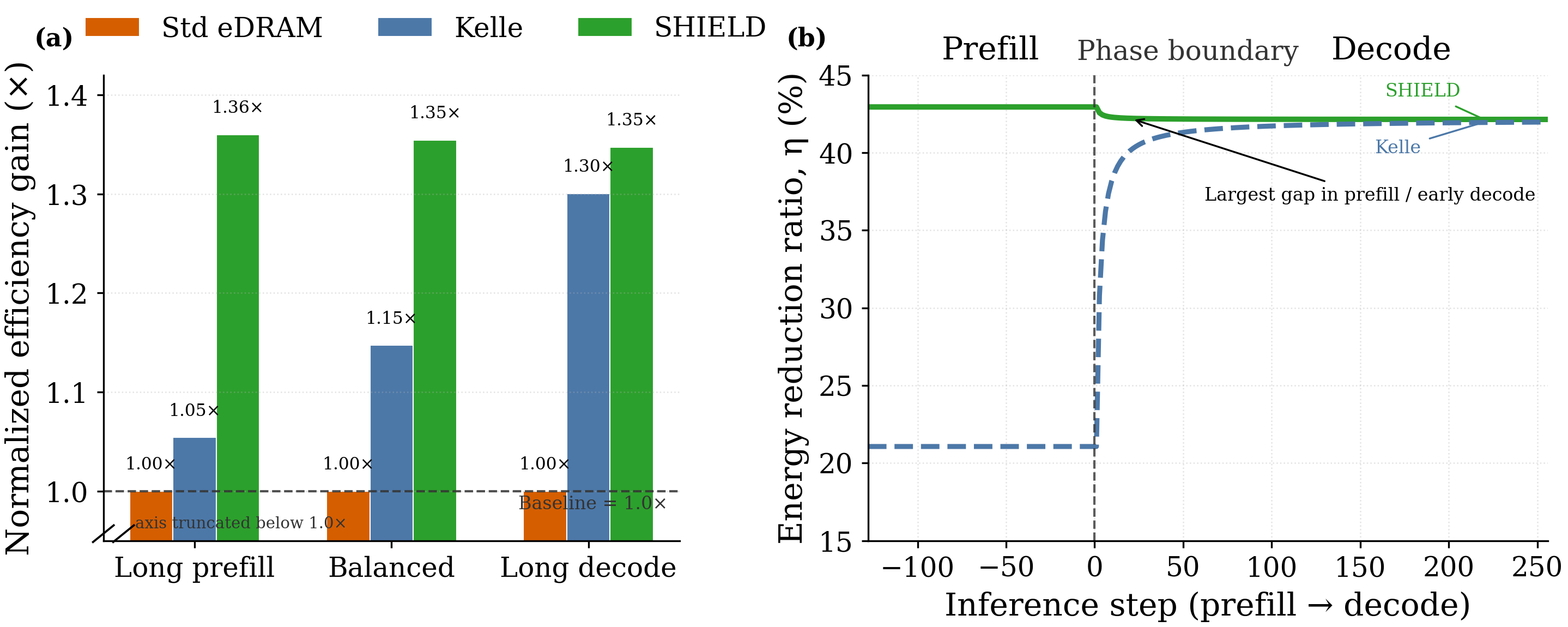}
    \caption{System-level energy evaluation of SHIELD. Left: efficiency gain across representative inference scenarios. Right: energy reduction ratio ($\eta$) over the inference lifecycle.}
    \label{fig:energy_eval_combined}
\end{figure}

\begin{table*}[t] 
\centering
\caption{Robustness Evaluation across Models}
\label{tab:robustness_full}
\addtolength{\tabcolsep}{-4pt} 
\resizebox{0.95\textwidth}{!}{
\begin{tabular}{l|ccc|ccc|ccc}
\toprule
\textbf{Model} & \multicolumn{3}{c|}{\textbf{WikiText-2 (PPL)} $\downarrow$} & \multicolumn{3}{c|}{\textbf{PIQA Accuracy} $\uparrow$} & \multicolumn{3}{c}{\textbf{ARC-Easy Accuracy} $\uparrow$} \\
 & Baseline & Kelle ($\Delta$) & \textbf{SHIELD ($\Delta$)} & Baseline & Kelle ($\Delta$) & \textbf{SHIELD ($\Delta$)} & Baseline & Kelle ($\Delta$) & \textbf{SHIELD ($\Delta$)} \\
\midrule
\textbf{Qwen3-1.7B} & 22.03 & +0.03 & \textbf{+0.02} & 72.25\% & -0.43\% & \textbf{0.00\%} & 72.77\% & -0.55\% & \textbf{-0.55\%} \\
\textbf{Qwen3-4B}   & 17.31 & +0.00 & \textbf{+0.02} & 75.19\% & -0.38\% & \textbf{+0.22\%} & 80.43\% & +0.04\% & \textbf{0.00\%} \\
\textbf{Qwen3-8B}   & 12.74 & +0.01 & \textbf{+0.00} & 76.61\% & -0.44\% & \textbf{-0.22\%} & 83.42\% & +0.12\% & \textbf{+0.08\%} \\
\textbf{Mistral-7B} & 8.67  & -0.00 & \textbf{0.00}  & 81.99\% & 0.00\%  & \textbf{+0.06\%} & 83.54\% & -0.21\% & \textbf{+0.05\%} \\
\textbf{Llama-3-8B} & 9.01  & +0.00 & \textbf{+0.00} & 79.98\% & -0.11\% & \textbf{+0.27\%} & 82.15\% & -0.08\% & \textbf{0.00\%} \\
\bottomrule
\end{tabular}}
\vspace{-10pt} 
\end{table*}

\section{Evaluation}

We evaluate SHIELD from four perspectives. First, we justify the eDRAM-based activation pool used in SHIELD. Second, we formulate the baseline and segmented refresh-power models and derive the resulting energy reduction. Third, we measure efficiency gains across representative workloads and over the prefill-to-decode lifecycle. Finally, we evaluate robustness across diverse LLMs under lifecycle-aware refresh gating.

\subsection{Density and Baseline Efficiency}

We first justify the use of an eDRAM-based activation pool using \textit{DESTINY}~\cite{7092634}. For a 2~MB workspace, a pure SRAM implementation consumes 452.25~mW of leakage power, whereas the eDRAM data array leaks only 0.95~mW. Even after accounting for standard refresh overhead, eDRAM remains more energy-efficient for the high-density storage required by edge LLM inference. This result motivates the eDRAM-based workspace assumed in SHIELD.

\subsection{Energy Efficiency}

We model the total power consumption of the activation workspace as the sum of leakage power and refresh power. The baseline power of a conventional eDRAM-based workspace is
\begin{IEEEeqnarray}{rCCCl}
P_{base} & = & P_{leak} + P_{ref\_std} & = & P_{leak} + \frac{E_{ref\_cycle}}{T_{std}}
\end{IEEEeqnarray}
where $P_{leak}$ is the array leakage power, $E_{ref\_cycle}$ is the energy consumed per refresh cycle, and $T_{std}=45\,\mu s$ is the standard refresh interval~\cite{4700556}.

SHIELD partitions the activation workspace into three banks according to numerical sensitivity and tensor lifecycle.
Let $B_{total}$ denote the total workspace capacity, with $B_{KV}$ and $B_{QO}$ corresponding to the KV cache and QO allocations, respectively.
The resulting power consumption is
\begin{IEEEeqnarray}{rCl}
P_{SHIELD} & = & P_{leak} + P_{ref,exp} + P_{ref,kv} + P_{ref,qo}.
\end{IEEEeqnarray}

The exponent and sign segment, $P_{ref,exp}$, corresponds to the 9 most significant bits in BF16 and is refreshed at the standard interval:
\begin{equation}
P_{ref,exp} = \frac{9}{16}\cdot\left(\frac{E_{ref\_cycle}}{T_{std}}\right).
\end{equation}

The KV cache mantissa segment, $P_{ref,kv}$, corresponds to the 7-bit mantissas of persistent KV cache. These banks use a relaxed refresh interval of $T_{rel}=1216\,\mu s$, giving
\begin{equation}
P_{ref,kv} = \left(\frac{7}{16}\cdot\frac{B_{KV}}{B_{total}}\right)\cdot\left(\frac{E_{ref\_cycle}}{T_{rel}}\right).
\end{equation}

The transient QO mantissa segment, $P_{ref,qo}$, corresponds to the 7-bit mantissas of QO tensors. Because QO is short-lived, these banks operate under a refresh-less policy:
\begin{equation}
P_{ref,qo} = \left(\frac{7}{16}\cdot\frac{B_{QO}}{B_{total}}\right)\cdot 0 = 0.
\end{equation}

The normalized energy reduction of SHIELD relative to the baseline is
\begin{IEEEeqnarray}{rCl}
\eta & = & 1 - \frac{P_{SHIELD}}{P_{base}} \approx 1 - \left[ \frac{9}{16} + \frac{7}{16}\left(\frac{B_{KV}}{B_{total}}\cdot\frac{T_{std}}{T_{rel}}\right) \right].
\end{IEEEeqnarray}

Thus, the achievable energy reduction depends on the footprint ratio $B_{KV}/B_{total}$. We therefore evaluate SHIELD at both workload and token-lifecycle granularity.

\subsection{Cross-Scenario Performance Analysis}

We evaluate the efficiency gain, $P_{base}/P_{SHIELD}$, across three representative workloads: \textit{Summary} (prefill-dominant), \textit{Translation} (balanced), and \textit{Storytelling} (decode-dominant). Fig.~\ref{fig:energy_eval_combined}(a) shows that Kelle exhibits workload-dependent efficiency, with gains ranging from $1.15\times$ to $1.32\times$ as the relative QO footprint changes. In contrast, SHIELD maintains a nearly constant gain of about $1.35\times$ across all three scenarios. The advantage is largest when transient QO dominates, but remains stable even as the KV cache becomes the main memory component.

We next evaluate $\eta$ over a representative generation trace (Prefill=128, Decode=256) to complement the workload-level results. Fig.~\ref{fig:energy_eval_combined}(b) shows that SHIELD’s advantage over Kelle is largest during prefill and early decode, when transient QO occupies a larger fraction of the active workspace. As decoding proceeds, the KV cache grows and the efficiency gap narrows. Even so, SHIELD remains the most efficient design because it continues to eliminate refresh for the active QO mantissa banks.

\subsection{Robustness Evaluation}

We evaluate SHIELD’s robustness by injecting lifecycle-aware faults into mantissa segments during the forward pass of five state-of-the-art LLMs. Table~\ref{tab:robustness_full} compares SHIELD against Kelle, which applies an overall $2\times10^{-3}$ BER policy to the KV cache.

Across WikiText-2~\cite{merity2016pointer}, PIQA~\cite{Bisk_Zellers_Lebras_Gao_Choi_2020}, and ARC-Easy~\cite{clark2018think}, SHIELD maintains near-zero deviation from the error-free baseline ($\Delta \approx 0$). At the same time, it achieves higher energy savings than KV-only policies without the perplexity degradation observed under more aggressive refresh relaxation. These results show that lifecycle-aware refresh gating provides a robust operating envelope across diverse LLM families.

\section{Conclusion}
This paper presented SHIELD, a lifecycle-aware eDRAM memory architecture with hierarchical refresh control for energy-efficient LLM inference on edge NPUs.
By exploiting the short residency of QO activations and the lower numerical sensitivity of BF16 mantissas, SHIELD removes refresh from transient QO mantissa storage and assigns a relaxed refresh policy to persistent KV mantissa storage.
Across diverse LLMs and inference scenarios, SHIELD achieves a 35\% reduction in eDRAM refresh energy with negligible accuracy impact on WikiText-2, PIQA, and ARC-Easy.
Future work will investigate support for additional data formats and tighter co-design between floating-point execution units and segmented memory hierarchies to further reduce bit-shuffling overhead and data movement.

\bibliographystyle{IEEEtran}
\bibliography{citations}

\vspace{12pt}

\end{document}